\pdfoutput=1
\documentclass[epj,article]{svjour}
\usepackage{graphicx}
\usepackage{amsmath}
\usepackage{times}
\newcommand{\n}{{\rm n}}
\newcommand{\Hg}{{\rm Hg}}

\begin{document}
\title{Measurement of a false electric dipole moment signal from $^{199}$Hg atoms exposed to an inhomogeneous magnetic field}

\author{
S.~Afach \inst{1,2,3},
C.~A.~Baker \inst{4},
G.~Ban \inst{5},
G.~Bison \inst{2},
K.~Bodek \inst{6}, 
Z.~Chowdhuri \inst{2},
M.~Daum \inst{2},       
M.~Fertl \inst{1,2}\thanks{Now at University of Washington, Seattle WA, USA.},		
B.~Franke \inst{1,2}\thanks{Now at Max-Planck-Institute of Quantum Optics, Garching, Germany.},
P.~Geltenbort \inst{7},
K.~Green \inst{4,8},
M.~G.~D.~van~der~Grinten \inst{4}, 
Z.~Grujic \inst{9},
P.~G.~Harris \inst{8},
W. Heil \inst{10} ,
V.~H\'elaine \inst{2,5}\thanks{Now at LPSC, Grenoble, France.},  	
R.~Henneck \inst{2},
M.~Horras \inst{1,2}\thanks{Present address: Hauptstrasse 60, CH-4455 Zunzgen, Switzerland.},
P.~Iaydjiev \inst{4}\thanks{On leave from INRNE, Sofia, Bulgaria.},
S.~N.~Ivanov \inst{4}\thanks{On leave from PNPI, St. Petersburg, Russia.},
M.~Kasprzak \inst{9},
Y.~Kerma\"{i}dic \inst{11},
K.~Kirch \inst{1,2},
P. Knowles \inst{9}\thanks{Present address: Rilkeplatz 8/9, A-1040 Vienna, Austria.},
H.-C.~Koch \inst{9,10} ,
S.~Komposch \inst{1,2},
A.~Kozela \inst{12},
J.~Krempel \inst{1},
B.~Lauss \inst{2},
T.~Lefort \inst{5},
Y.~Lemi\`ere \inst{5},
A.~Mtchedlishvili \inst{2},     
O.~Naviliat-Cuncic \inst{5}\thanks{Now at Michigan State University, East-Lansing, USA.},
J.~M.~Pendlebury \inst{8},
F.~M.~Piegsa \inst{1},
G.~Pignol \inst{11},
P.~N.~Prashant \inst{2,13},
G.~Qu\'em\'ener \inst{5},
D.~Rebreyend \inst{11},
D.~Ries \inst{2},
S.~Roccia \inst{14},
P.~Schmidt-Wellenburg \inst{2},
N.~Severijns \inst{13},
A. Weis \inst{9},
E.~Wursten \inst{13},
G.~Wyszynski \inst{1,6}, 	
J.~Zejma \inst{6},
J.~Zenner \inst{1,2},
G.~Zsigmond \inst{2} 
}                     
\institute{
ETH Z\"urich, Institute for Particle Physics, CH-8093 Z\"urich, Switzerland 
\and
Paul Scherrer Institute (PSI), CH--5232 Villigen-PSI, Switzerland
\and
Hans Berger Department of Neurology, Jena University Hospital, D-07747 Jena, Germany
\and
Rutherford Appleton Laboratory, Chilton, Didcot, Oxon OX11 0QX, United Kingdom
\and
LPC Caen, ENSICAEN, Universit\'e de Caen Basse-Normandie, CNRS/IN2P3, Caen, France
\and
Marian Smoluchowski Institute of Physics, Jagiellonian University, 30--059 Cracow, Poland
\and
Institut Laue--Langevin, Grenoble, France
\and
Department of Physics and Astronomy, University of Sussex, Falmer, Brighton BN1 9QH, United Kingdom
\and
Physics Department, University of Fribourg, CH-1700 Fribourg, Switzerland
\and
Institut f\"ur Physik, Johannes--Gutenberg--Universit\"at, D--55128 Mainz, Germany
\and
LPSC, Universit\'e  Grenoble Alpes, CNRS/IN2P3, Grenoble, France
\and
Henryk Niedwodnicza\'nski Institute for Nuclear Physics, 31--342 Cracow, Poland
\and
Instituut voor Kern-- en Stralingsfysica, University of Leuven, B--3001 Leuven, Belgium
\and
CSNSM, Universit\'e Paris Sud, CNRS/IN2P3, Orsay Campus, France
}
\mail{kermaidic@lpsc.in2p3.fr,rebreyend@lpsc.in2p3.fr}
\date{Received: date / Revised version: date}
\abstract{
 We report on the measurement of a Larmor frequency shift proportional to the electric-field strength for $^{199}{\rm Hg}$ atoms contained in a volume permeated with aligned magnetic and electric fields. This shift arises from the interplay between the inevitable magnetic field gradients and the motional magnetic field. The proportionality to electric-field strength makes it apparently similar to an electric dipole moment (EDM) signal, although unlike an EDM this effect is P- and T-conserving. 
We have used a neutron magnetic resonance EDM spectrometer, featuring a mercury co-magnetometer and an array of external cesium magnetometers, to measure the shift as a function of the applied magnetic field gradient.  Our results are in good agreement with theoretical expectations.
%
}
\authorrunning
\titlerunning
\maketitle
\section{Introduction}
\label{sec:intro}
Recent investigations characterizing frequency shifts for spins contained in vessels permeated with magnetic and electric fields $B$, $E$ have been motivated principally by the search for electric dipole moments (EDMs) of simple non-degenerate systems (neutron, atoms, molecules) and the potential discovery of new sources of CP violation \cite{ram2013}. Such experiments look for shifts, proportional to an applied electric field, of the Larmor precession frequency of stored particles.  
Any additional such shift is therefore a potential source of systematic errors. Among the few magnetic-field related spurious shifts, 
one is of particular concern: due to 
the motional magnetic field ${\mathbf E \times \mathbf v/{c^2} }$, a shift arises that is proportional to the electric-field strength and therefore mimics an EDM signal.
Interestingly enough, ${\mathbf E \times \mathbf v/{c^2} }$ effects were already the main limiting factor for the early neutron beam experiments \cite{ramsey1982}. Then, with the advent of the storable ultra-cold neutrons (UCN), it was erroneously assumed for many years that this false EDM signal would vanish, based on the argument that the velocity of trapped particles averages to zero. The first correct and comprehensive calculation of this effect was given in Ref.~\cite{pendlebury2004}, in the context of an EDM experiment with stored particles. 
For completeness, it should be mentioned that 
Stark interference effects, such as the one reported for $^{199}{\rm Hg}$ in Ref.~\cite{loftus2011}, are also known to produce false EDM signals for atoms. The effect discussed in the present article is of a different nature, and to make the distinction we will refer to it as the motional false EDM. 
%
%

Our collaboration is conducting a program to search for the neutron EDM \cite{baker2011}, using the new ultracold neutron (UCN) source \cite{Lauss2014} at the Paul Scherrer Institute (PSI). We are currently working with an upgraded version of the spectrometer \cite{baker2013} that was used to establish the best nEDM limit, 
$$\left| d_\n \right| < 2.9 \times 10^{-26}\, e\,\text{cm} \, (90\% \, \text{C.L.}),$$ 
at the Laue Langevin Institute (ILL) \cite{baker2006}. One distinct feature of this device is a mercury co-magnetometer \cite{green1998} using a spin-polarized vapor of $^{199}$Hg atoms that precess in the same volume as the neutrons. The nEDM analysis is then based on the ratio of the Larmor precession frequencies, $R=f_\n/f_{\Hg}$, which to first order is free of magnetic field fluctuations. 
However, both neutrons and mercury atoms are subject to a frequency shift that is proportional to the electric field, due to the unavoidable presence of magnetic-field gradients.
As will be shown, the motional false neutron EDM, $d_\n^{\rm false}$, is negligible, at least at the current level of sensitivity. In contrast, the mercury-induced false nEDM 
\begin{equation}
d_\n^{{\rm false}, \Hg} = \frac{\gamma_\n}{\gamma_{\Hg}} d_{\Hg}^{\rm false}	\approx 3.8\, d_{\Hg}^{\rm false},
\end{equation}
where $d_{\Hg}^{\rm false}$ is the motional mercury false EDM and $\gamma_\n$, $\gamma_{\Hg}$ are the gyromagnetic ratios of the neutron and $^{199}$Hg respectively, is a major systematic effect that must be precisely controlled. 

One of the main improvements accomplished recently within the experiment is the installation of an array of cesium magnetometers that surrounds the precession chamber. This new device has made it possible to measure the magnetic field distribution, and thus to calculate the vertical gradient in the trap, which underlies the false EDM discussed here. 

In this article, we report on the first direct measurement of a motional false EDM signal for stored mercury atoms. A comparison to theoretical expectations is also presented.

\section{Theory of frequency shifts induced by magnetic field gradients: a brief reminder}
\label{sec:theory}

Particles with a magnetic moment exposed to a magnetic field, ${\bf B_\textnormal{0}} = B_0 {\bf \hat{z}}$, precess at the Larmor frequency $f_{{\rm L}} = \gamma \, B_0 / 2 \pi$ where $\gamma$ is the gyromagnetic ratio. Because of experimentally unavoidable magnetic field gradients, the Larmor frequency of a particle moving through this field will be subject to a shift, known as the Ramsey-Bloch-Siegert (RBS) shift \cite{ramsey1955}. If an electric field ${\bf E}$ (parallel or anti-parallel to ${\bf B_\textnormal{0}}$) is applied -- as is the case in experiments searching for EDMs -- the moving particle will experience an additional motional magnetic field ${\mathbf B_v = \mathbf E \times \mathbf v/{c^2} }$. It is the interplay between this field and the magnetic field gradients that lies at the origin of a frequency shift proportional to the electric field strength, thus inducing a false EDM.

As mentioned above, the first detailed calculation of such false EDMs for stored particles was given in Ref.~\cite{pendlebury2004} in the context of the RAL-Sussex-ILL neutron EDM experiment \cite{baker2006}. The authors derived expressions for the two limiting cases: non adiabatic and adiabatic, corresponding to $2\pi f_{\rm L} \tau \gg 1$ and $2\pi f_{\rm L} \tau  \ll 1$ respectively, where $\tau$ is the typical time particles take to cross the trap. Both regimes are of interest, since $^{199}{\rm Hg}$ atoms fall into the first category whereas UCNs fall into the second. 
More general results, valid for a broad range of frequencies, were obtained only for cylindrical symmetry and specular reflections. The expressions of the frequency shifts for the two limiting regimes are :

\begin{align}
 		\delta f_\textrm{L}  &= \frac{\gamma^2 D^2}{32 \pi \, c^2} \frac{\partial B_0}{\partial z} E & \quad \textrm{(non adiabatic)} 
		\label{eq_deltaOmegaNonAdiabatic}\\
 		\delta f_\textrm{L} &= \frac{v_{xy}^2}{4\pi\, B_0^2\, c^2} \frac{\partial B_0}{\partial z} E & \quad \textrm{(adiabatic),}
 		\label{eq_deltaOmegaAdiabatic}
\end{align}

\noindent where $\gamma$ is the gyromagnetic ratio, $D$ is the diameter of the trap, $c$ is the velocity of light and $v_{xy}$ is the particle velocity transverse to $B_0$. Note the absence of the gyromagnetic ratio in Eq.~(\ref{eq_deltaOmegaAdiabatic}). Indeed, in the adiabatic case, the frequency shift can be interpreted as originating from a phase of purely geometric nature, or Berry's phase \cite{ber1984,commins1991}, and is therefore independent of the coupling strength to the magnetic field.

These results were then complemented and extended using the general theory of relaxation (Redfield theory) \cite{lamoreaux2005,pignol2012}, and then by solving the Schr\"odinger equation directly \cite{steyerl2014}. In Ref.~\cite{pignol2012}, an expression valid for arbitrary field distributions or trap shapes was obtained in the non-adiabatic limit :

\begin{align}
 		\delta f_\textrm{L} &= \frac{\gamma^2}{2 \pi c^2} \left\langle x B_x\, + yB_y \right\rangle E   \quad \textrm{(non adiabatic),}									
		\label{eq_deltaOmegaNonAdiabaticGeneralized}
\end{align}
where the brackets refer to the average over the storage volume. For a cylindrical uniform gradient and a trap with cylindrical symmetry, Eq.~(\ref{eq_deltaOmegaNonAdiabaticGeneralized}) reduces to Eq.~(\ref{eq_deltaOmegaNonAdiabatic}).

Using the relationship between the frequency shift and the false EDM, 

$${d}^{\rm false} = \frac{h}{2E} \delta f_{\rm L} (E)$$

\noindent where $h$ is Planck's constant, together with 
Eqs.~(\ref{eq_deltaOmegaNonAdiabatic}) and (\ref{eq_deltaOmegaAdiabatic}), one can now readily calculate the magnitude of the false EDMs for the mercury and for the neutron (both direct and mercury induced).  Given our experimental conditions (see section~\ref{sec:setup}) and assuming a neutron velocity of 3 m/s, one obtains: 
\begin{align}
 		&d_\n^{\rm false} = \frac{\partial B_0}{\partial z} \, 1.490 \times 10^{-29}\, e \, \text{cm}/\text{(pT/cm)}   \\
 		& \nonumber\\
 		&d_{\Hg}^{\rm false} = \frac{\partial B_0}{\partial z} \, 1.148 \times 10^{-27}\, e \, \text{cm}/ \text{(pT/cm)} \label{falseHgEDMTheory}\\
 		& \nonumber \\
 		&d_\n^{\rm false, \Hg} = \frac{\partial B_0}{\partial z} \, 4.418 \times 10^{-27}\, e \, \text{cm}/\text{(pT/cm).}
\end{align}
Considering a typical value of 10 pT/cm for the vertical ($z$ direction) gradient in our setup, we can conclude on the one hand that the direct false neutron EDM is negligible, at least at the current level of sensitivity. On the other hand, the mercury-induced false neutron EDM is a major systematic error that must be properly taken into account. 
\section{Experimental apparatus}
\label{sec:setup}

The experimental study was performed with the nEDM spectrometer installed at the PSI UCN source. This room-temperature apparatus uses the Ramsey method of separated oscillatory fields \cite{green1998,ramsey1950} to search for a shift, proportional to the strength of an applied electric field, in the neutron Larmor precession frequency.

\begin{figure}[ht]
	\centering
	\includegraphics[width=\linewidth]{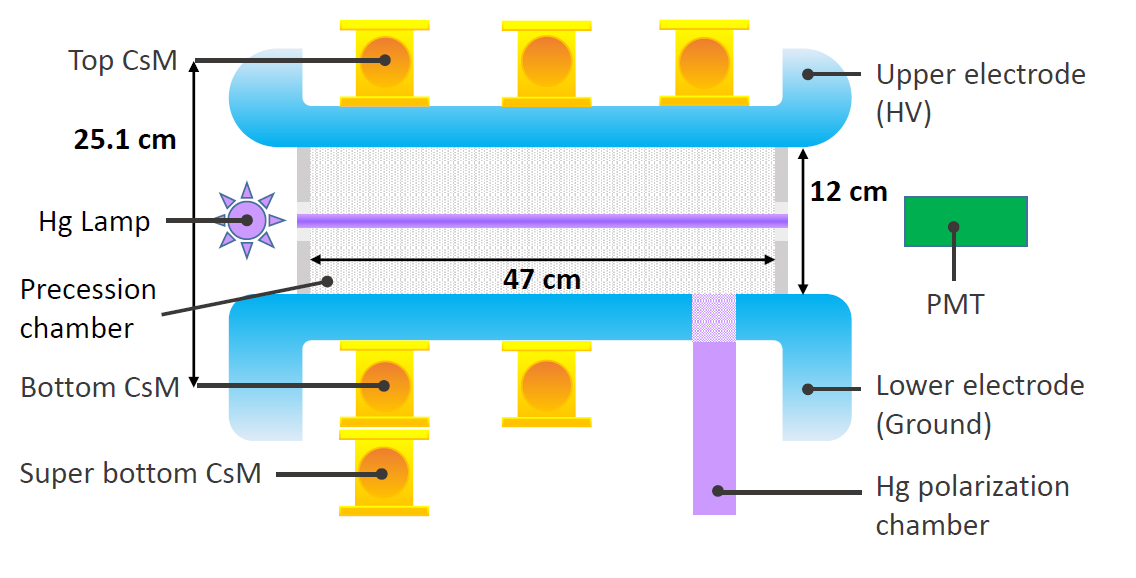}		
	\caption{Schematic view of the precession chamber of the nEDM@PSI experiment.}
	\label{fig:oILL}      
\end{figure}

Under normal operation, polarized UCNs are stored in a $\sim20$~liter chamber (internal diameter {\it D}~=~47~cm, height {\it H}~=~12~cm), composed of a hollow polystyrene cylinder (coated with deuterated polystyrene) \cite{bodek2008,kuzniak2008} and two disk-shaped aluminum electrodes coated with diamond-like carbon (Fig.~\ref{fig:oILL}). A cos$\theta$ coil produces a highly homogeneous magnetic field, $B_0 \approx 1 \, \mu\text{T}$, in the vertical direction while  the two electrodes -- the top one being connected to a high voltage (HV) source and the bottom one to ground potential -- generate a strong electric field ($E \approx 10\, \text{kV/cm}$), either parallel or anti-parallel to  ${\bf B_\textnormal{0}
}$. In addition, a set of trim coils permits an optimization of the magnetic field uniformity at the $10^{-3}$ level.

The key to such experiments relies on the ability to control the magnetic field both in terms of stability and homogeneity. To this end, we use two highly sensitive and complementary atomic magnetometers based on mercury ($^{199}{\rm Hg}$) and cesium ($^{133}$Cs) atoms, respectively. Mercury is used in a co-magnetometer mode: polarized mercury atoms precess in the same volume as the neutrons, hence probing approximately the same space- and time-averaged magnetic field.  Cesium is used in a set of external magnetometers surrounding the storage chamber. The former is an ideal tool to correct for field drifts, while the latter gives access to the spatial field distribution. 

\subsection{The mercury co-magnetometer}

To date, $^{199}{\rm Hg}$ is the only atomic element that has been used as a co-magnetometer for a neutron EDM experiment. Thanks to its nuclear polarization, it benefits from long wall collision relaxation times, and polarization lifetimes larger than 100~s can be achieved. 
Moreover, it is one of the rare elements in which nuclear spin polarization can be created and monitored by optical means. 
It is worth noting that the best absolute EDM limit comes from an experiment using $^{199}{\rm Hg}$\cite{griffith2009}\footnote{One may wonder why the effect discussed in the present article was not observed in that experiment. They actually use spectroscopy cells filled with 475 Torr of CO buffer gas acting as a UV quencher. Consequently, mercury atoms move in the diffusive regime where the motional false EDM essentially vanishes -- in contrast to the ballistic regime of our mercury co-magnetometer.}: 

$$ \left| d({\rm ^{199}Hg}) \right| < 3.1 \times 10^{-29}\, e\,\text{cm} \, (95\% \, {\rm CL}).$$ 

In our experiment, a vapor of mercury atoms is spin-polarized by optical pumping in a polarization chamber located underneath the precession chamber (Fig.\ \ref{fig:oILL}). The operation of the co-magnetometer is synchronous with the nEDM measurement, and follows cycles about 300~s long. During neutron counting and filling, mercury atoms are continuously injected and optically pumped in the polarization chamber. Once the precession chamber is filled with UCNs, we let the vapor diffuse into the precession chamber where, after the application of a $\pi / 2$ pulse, the atoms freely precess around  ${\bf B_\textnormal{0}}$ at a frequency of about 8~Hz.
The interaction of the precessing atoms with a circularly polarized resonant probe beam produces a light-intensity modulation whose analysis yields the Larmor frequency of the atoms. 

\begin{figure}[h]
	\centering
	\includegraphics[width=\linewidth]{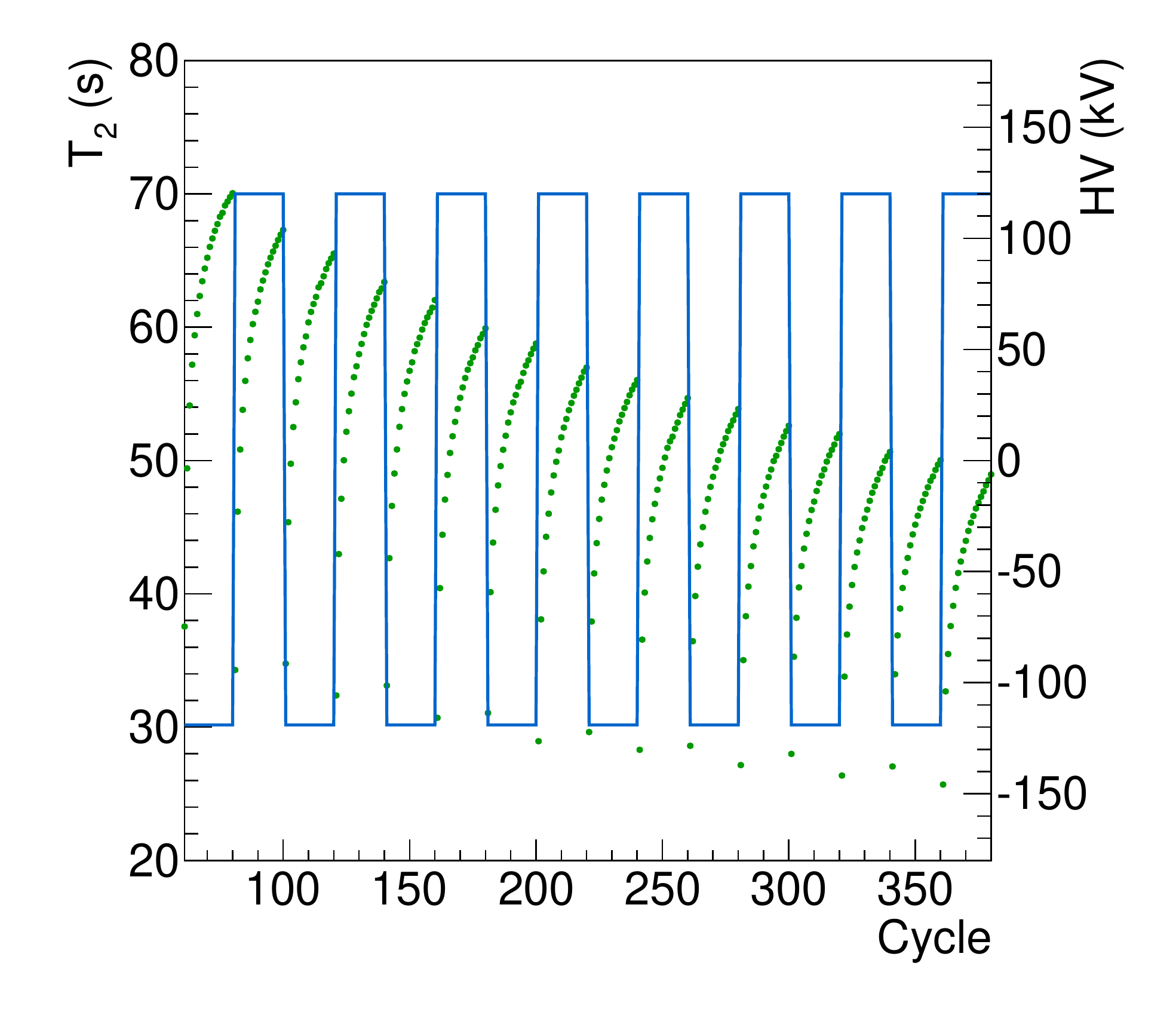}		
	\caption{Transverse relaxation time  T$_2$ of $^{199}\text{Hg}$ atoms (green points) together with the high voltage value (blue line) versus cycle number. Sudden drops of T$_2$ are observed after each polarity reversal.}
	\label{fig:tauHg}      
\end{figure}

One of the major drawbacks of the $^{199}{\rm Hg}$ co-magnetometer is its sensitivity to high voltage. As illustrated in Fig.\ \ref{fig:tauHg}, which displays the transverse polarization relaxation time T$_2$ versus the cycle number, sudden T$_2$ drops are systematically observed after each HV polarity reversal. The corresponding reduction of the signal amplitude directly affects the precision of the magnetometer. Fortunately, optimal performance can be recovered via  discharge cleaning in an oxygen atmosphere. On average, the precision of the mercury co-magnetometer is of the order of 100~fT, equivalent to a magnetometric precision at the 0.1~ppm level per cycle.

\subsection{The array of cesium magnetometers}
\label{sec:CsM}
\begin{figure}[h]
	\centering
	\includegraphics[width=\linewidth]{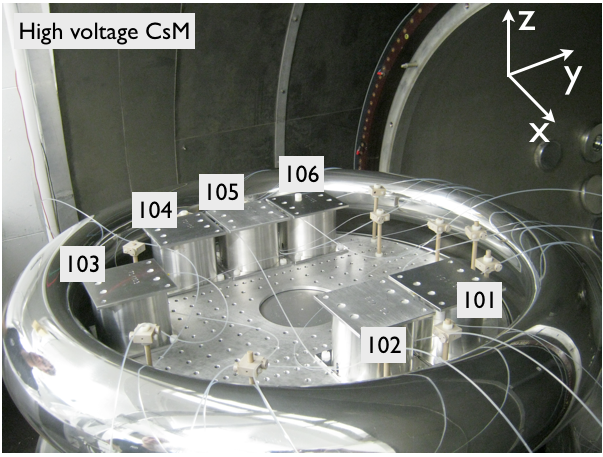}		
	\caption{Picture of the six HV-compatible Cs magnetometers installed on the top HV electrode in Al enclosures. Optical fibers are also visible.}
	\label{fig:HV-CsM}      
\end{figure}

An array of 16 cesium magnetometers (CsM) \cite{Knowles2009} allows measurement of the magnetic field distribution in the region of interest and, in particular, it gives us knowledge of the vertical gradient $\partial B_0 / \partial z$. Six HV-compatible (i.e.\ fully optically coupled) magnetometers were placed on top of the precession chamber, and ten standard ones below (Fig.~\ref{fig:oILL} and \ref{fig:HV-CsM}). These laser-pumped magnetometers use a vapor of $^{133}$Cs atoms (gyromagnetic ratio $\gamma = 2 \pi \times 3.5\, \text{kHz}/\mu \text{T}$) and are operated in a phase-stabilized mode. 
They have a high statistical sensitivity ($\sim$ 100~fT for 40~s long measurements); however, they suffer from inaccurracies of their absolute field readings, with offsets that can be as high as 100 pT. They are therefore precise but not accurate.
Finally, it is important to note that these magnetometers -- like the mercury co-magnetometer -- are scalar: they measure the magnitude of the magnetic field at the center of the bulb containing the cesium vapor.

\section{Measurement and data analysis}
\label{sec:analysis}

A preliminary measurement with a limited number of CsM was performed in 2011, and led to a first result \cite{marlonThesis}. The present analysis is based on a dedicated data-taking period of 2 weeks' duration in December 2013, where eight different gradient settings were explored: four with the magnetic field pointing upwards ($B_0^{\uparrow}$), and four downwards ($B_0^{\downarrow}$). Two trim coils were used to set a vertical gradient in addition to the $B_0$ field generated by the main coil. For each field configuration, about 500 cycles were recorded with a basic HV polarity pattern $(+\,-\,-\,+)$ and polarity changes every 20 cycles. The voltage was set to 120~kV, i.e. as high as possible to maximize the frequency shift while preserving a smooth operation (limited number of electrical breakdowns). 

As discussed above, the frequent polarity reversals induced a significant degradation of the mercury magnetometer's sensitivity. Consequently, we decided to limit the free precession time to 40~s, a good compromise between sensitivity and the number of cycles. The mercury frequency was extracted using our standard ``two windows'' method \cite{chibane1995}. It consists of fitting the signal phase at either end of the signal, using data in two 15~s windows at the beginning and end of the time series. 
This method optimally takes into account  possible frequency drifts during the precession time. 
During data taking, the mercury frequency uncertainty varied in the range 1-2~$\mu \text{Hz}$.

Outputs from all 16 CsM were continuously recorded at a rate of 1~Hz, and a mean value of the magnetic field was calculated for time periods having an exact overlap with the mercury precession. We further made the approximation 

$$B_{\rm CsM} = \sqrt[2]{B_z^2 + B_T^2} \approx B_z(\vec{r}_{\rm CsM}),$$

\noindent where $B_T$ is a small transverse component. From several 3D mapping campaigns during which all coils (main and trim) were mapped, we know that this approximation is valid at the 10$^{-4}$ level.

\subsection{Gradient extraction}
\label{subsec:gzExtrac}

We extracted the vertical gradient by fitting a harmonic polynomial expansion of the magnetic field to the CsM array data. The choice of harmonic polynomials ensures that the resulting expressions satisfy Maxwell's equations. 
Due to the limited number of magnetometers the expansion was limited to the next-to-linear order (NLO), which involves 9 parameters:
	\begin{align}
		B_z(x,y,z) =\, &b_0 + g_x \, x + g_y \, y + g_z \, z + \nonumber \\
									&g_{xx} (x^2 - z^2) + g_{yy}  (y^2 - z^2)+ \nonumber \\
									&g_{xy}  xy + g_{xz}  xz + g_{yz}  yz.
	\label{eqn:fit_NLO}
	\end{align}
From expression~(\ref{eqn:fit_NLO}), one can easily calculate the volume average of $B_z$ and of its vertical gradient, assuming a trap with cylindrical symmetry:
\begin{align}
B_0 \equiv \left\langle B_z \right\rangle &= b_0 +(g_{xx} + g_{yy})\left(\frac{D^2}{16}-\frac{H^2}{6}\right) \\
\left\langle \frac{\partial B_z}{\partial z}\right\rangle &= g_z.  
\label{eq:B_0}
\end{align}

Let us now turn to the delicate task of estimating gradient uncertainties. The two main sources that have to be taken into account are the error on the magnetic field, and the extraction procedure. To assess their respective effects, extensive studies have been carried out using a toy model to generate known field distributions and check the extracted parameters \cite{victorThesis}. 
It was found that the errors coming both from the magnetometer offsets and from the expansion truncation never exceed 5~pT/cm.
In addition, we used a technique known as the jackknife method to get an error directly from the data. It involves performing a series of  $\chi^2$ minimizations (unweighted in our case) by removing one out of the 16 magnetometers at a time. The dispersion of the extracted parameters provides an estimate of the error. For the different field configurations, we systematically obtained errors in the range $10 \pm 5 \, \text{pT/cm}$, consistent with the model outcome. These jackknife errors were used subsequently in the analysis. 

\subsection{Frequency shift measurement}
\label{subsec:freqShift}

\begin{figure}[h]
	\centering
	\includegraphics[width=\linewidth]{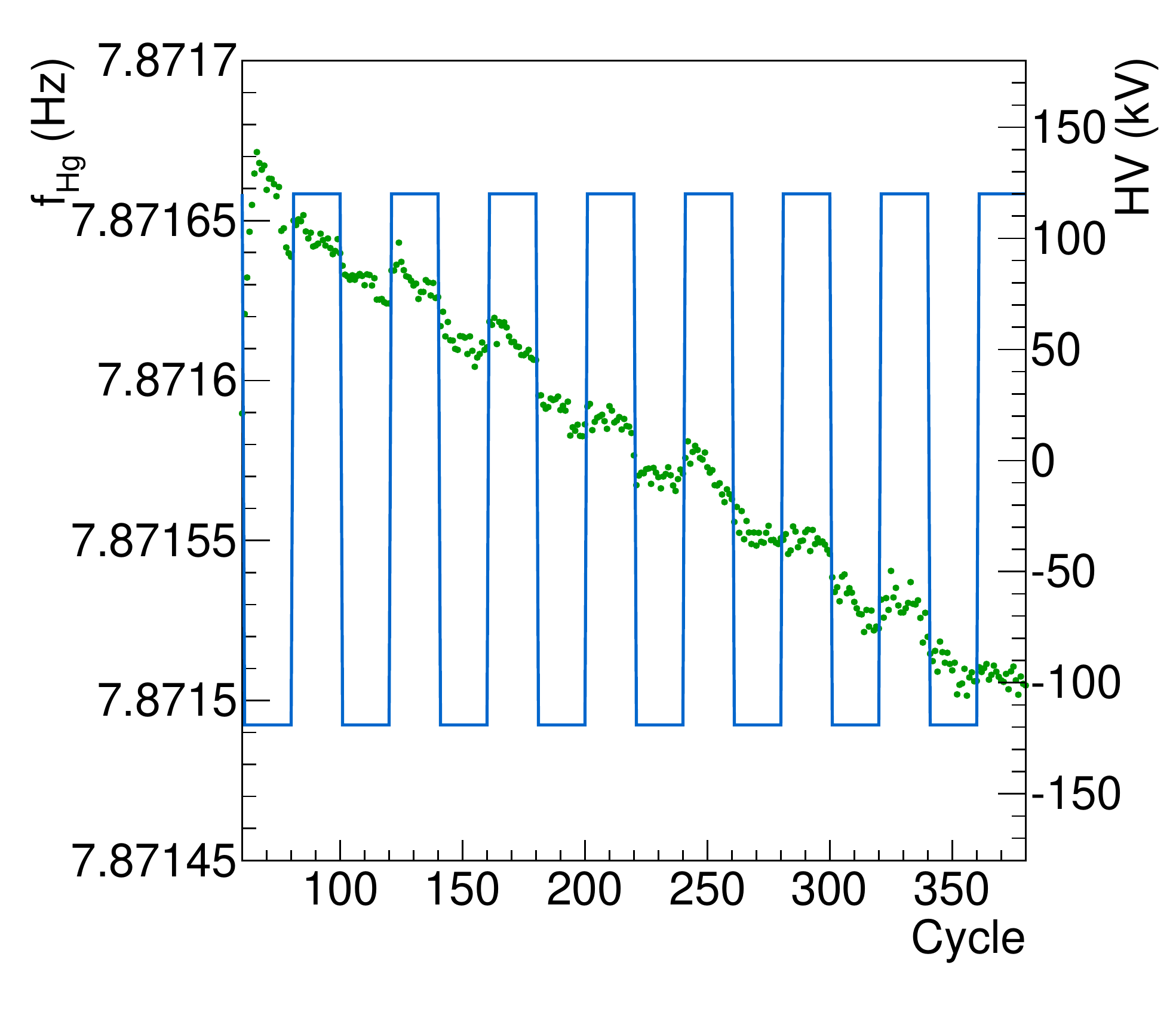}		
	\caption{Mercury frequency versus cycle number. The blue line shows the value of the applied high voltage.}
	\label{fig:fHg_raw}      
\end{figure}

A sample of a raw data time series $f_\Hg$ against cycle number is displayed in Fig.~\ref{fig:fHg_raw}, together with the corresponding high-voltage values. Despite the large point-to-point fluctuations and a slow linear drift, one can clearly observe a small but systematic correlation of the frequency shift with the electric field polarity. To correct for the slow magnetic field drift, we sliced the data relative to the electric field polarity and analyzed data sets corresponding to the $(+\,-\,-\,+)$ HV pattern. By doing so, any linear drift is exactly cancelled and higher orders are attenuated. 

For a data slice $(+\,-\,-\,+)$, corresponding to 40 cycles, the extracted frequency shift and its uncertainty are given by 

\begin{equation}
	\delta f_{\Hg} = \left\langle f_{\Hg}^+ \right\rangle - \left\langle f_{\Hg}^- \right\rangle
\end{equation}
\noindent and
\begin{equation}
	\Delta\delta f_{\Hg} = \sqrt{\Delta\left\langle f_{\Hg}^+ \right\rangle^2 + \Delta\left\langle f_{\Hg}^- \right\rangle^2},
\end{equation}

\noindent
where $\left\langle f_{\Hg}^{+(-)} \right\rangle$ and $\Delta\langle f_{\Hg}^{+(-)} \rangle$ stand for the mean frequency and its uncertainty as derived from the frequency distribution for the given HV polarity ($+$ or $-$).
Finally, a weighted mean over the whole set of data slices was performed to estimate the electric-field induced frequency shift $\delta f_{\rm L} (E)$ for a given vertical gradient.

\section{Results and discussion}
\label{results}

\begin{figure}[h]
	\centering
	\includegraphics[width=\linewidth]{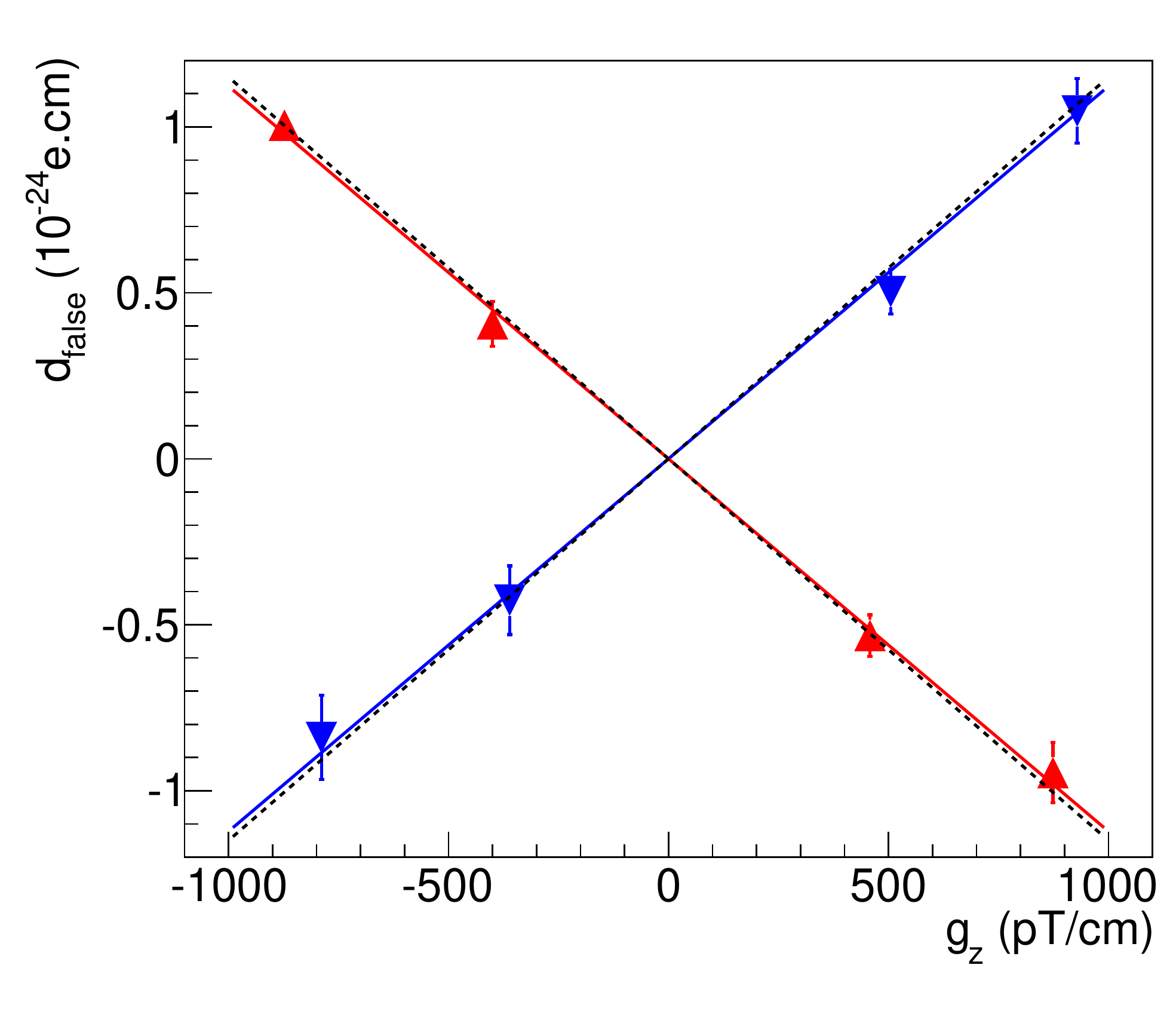}		
	\caption{Motional false mercury EDM versus the vertical gradient $g_z$ for $B_0^{\uparrow}$ (red up triangles) and $B_0^{\downarrow}$ (blue down triangles). The solid lines correspond to a linear fit, and the dashed line to the theory discussed in section \ref{sec:theory}. The horizontal error bars are smaller than the symbol size. }
	\label{fig:dFalseVSgz}      
\end{figure}

The final result is displayed in Fig.\ \ref{fig:dFalseVSgz}. The motional false mercury EDM 
is plotted against the extracted vertical gradient $g_z$. 
The solid lines (red for $B_0^{\uparrow}$, blue for $B_0^{\downarrow}$) correspond to a global linear fit with a single free parameter, namely the slope $a$ ($\chi^2/\nu = 2.1/7$).

We can now compare the measured slope to its theoretical expectation  from Eq.~6: 
\begin{align}
	\left| a_{\rm exp} \right| &= 1.122(35) \times 10^{-27} \, e \,\text{cm}/(\text{pT/cm}),\\
	\mathrm{and} \left| a_{\rm th} \right| &= 1.148 \times 10^{-27} \, e \,\text{cm}/(\text{pT/cm}).
\end{align}

The agreement at the 1$\sigma$ level makes us confident that our magnetic gradient extraction procedure is reliable.
This encouraging result is nonetheless not sufficient to directly control the mercury-induced false neutron EDM at the required level of sensitivity. Indeed, for an error of 10~pT/cm on the vertical gradient, Eq.~(\ref{eq_deltaOmegaNonAdiabatic}) translates to a systematic error of $4.4 \times 10^{-26} e \,\text{cm}$ on the neutron EDM, which is already larger than the current limit.
%
%
There is fortunately a way to circumvent this issue. In their last nEDM paper~\cite{baker2006}, the authors describe an analysis technique that enables one to find experimentally the working point with no vertical gradient and therefore no motional false EDM. 
This method, based on a tiny center of mass offset between the cold neutrons and the warmer mercury atoms, nevertheless induces some additional systematic errors. These errors were carefully assessed and found subdominant with a final result statistically limited.

Whereas the use of the Hg comagnetometer is essential and does not limit our nEDM sensitivity for the time being, with a foreseen sensitivity of a few $10^{-27} e \,\text{cm}$ in the coming years, new magnetometry solutions will be needed in the future. We pursue an intensive R\&D program on magnetometry using $^{133}$Cs but also $^3$He atoms \cite{koch2015}. In particular, efforts towards improving the absolute accuracy of the Cs magnetometers are currently underway \cite{grujic2015} as well as the implementation of Cs vector magnetometers \cite{afach2015}. In parallel,
we have started design and construction of a next generation nEDM spectrometer \cite{baker2011} which, among other improvements, will benefit from a much better magnetic field control (passive and active). Advanced magnetometry and improved magnetic shielding will be combined with co-magnetometry or could even allow operation with only external magnetometers. Any possible mercury-induced false motional nEDM will therefore be much further suppressed or completely avoided. 

\section{Conclusions}
\label{sec:conclusions}

We have performed a measurement of a frequency shift proportional to the electric field strength for stored $^{199}$Hg atoms\footnote{It should be noted that, strictly speaking, we have only observed that frequency shifts were E-odd (measurements were done at different gradients but at a single HV value). However, the absence of physical justification for higher-order odd terms together with the excellent agreement with theory led us to disregard this possibility.},
 using a spectrometer devoted to the search for the neutron electric dipole moment at PSI. This shift, which we call the motional false EDM, originates from the combination of vertical magnetic field gradients with the motional magnetic field and could be measured for the first time thanks to the unique combination of a mercury co-magnetometer and an array of external cesium magnetometers. 
The agreement with a prediction based on the general Redfield theory of relaxation provides additional confidence in the validity of our gradient-extraction procedure as well as in our capability to measure and control the vertical gradient. The same method was used in a recent measurement of the neutron magnetic moment \cite{afach2014}.

\section*{Acknowledgements}

We are grateful to the PSI staff (the accelerator operating team and the BSQ group) for providing excellent running conditions, and we acknowledge the outstanding support of M.~Meier and F.~Burri. Support by the Swiss National Science Foundation Projects 200020-144473 (PSI), 200021-126562 (PSI), 200020-149211 (ETH) and 200020-140421 (Fribourg) is gratefully
acknowledged. The LPC Caen and the LPSC acknowledge the support of the French Agence Nationale de la Recherche (ANR) under reference ANR-09–BLAN-0046. The Polish partners acknowledge The National Science Centre, Poland, for the grant No. UMO-2012/04/M/ST2/00556. This work was partly supported by the Fund for Scientific Research Flanders (FWO), and Project GOA/2010/10 of the KU Leuven. The original apparatus was funded by grants from the UK’s PPARC (now STFC).

\end{document}